# Magnetization in electron- and Mn- doped SrTiO$_3$

D. Choudhury[1,2], B. Pal[1], A. Sharma[2], S. V. Bhat[2], D. D. Sarma[1,3,4,5,*]

[1]Solid State and Structural Chemistry Unit, Indian Institute of Science, Bangalore- 560012, India

[2]Department of Physics, Indian Institute of Science, Bangalore- 560012, India

[3]Department of Physics and Astronomy, Uppsala University, Box 516, SE-75120, Uppsala, Sweden

[4]Council of Scientific and Industrial Research-Network of Institutes for Solar Energy (CSIR-NISE), New-Delhi, India

[5]Jawaharlal Nehru Center for Advanced Scientific Research, Bangalore-560054, India

* Correspondence to [sarma@sscu.iisc.ernet.in]

Mn-doped SrTiO$_{3.0}$, when synthesized free of impurities, is a paramagnetic insulator with interesting dielectric properties. Since delocalized charge carriers are known to promote ferromagnetism in a large number of systems via diverse mechanisms, we have looked for the possibility of any intrinsic, spontaneous magnetization by simultaneous doping of Mn ions and electrons into SrTiO$_3$ via oxygen vacancies, thereby forming SrTi$_{1-x}$Mn$_x$O$_{3-\delta}$, to the extent of making the doped system metallic. We find an absence of any enhancement of the magnetization in the metallic sample when compared with a similarly prepared Mn doped, however, insulating sample. Our results, thus, are not in agreement with a recent observation of a weak ferromagnetism in metallic Mn doped SrTiO$_3$ system.

SrTiO$_3$ crystallizes in the simplest possible cubic structure. It is a diamagnetic, high band-gap insulator and exhibits many technologically important properties, making it a promising material for various dielectric capacitor and DRAM applications [1,2]. Lots of effort have been devoted in making SrTiO$_3$ a dilute magnetic semiconductor material by suitably doping various transition-metal ions into it [3-5]. Recently, Mn doped SrTiO$_3$ has attracted increased attention due to claims of simultaneous presence of dielectric and magnetic glassiness [6]. Our earlier work on Mn doped SrTiO$_3$, however, has shown that very small amounts of Mn-oxide impurity phase, which is often overlooked in commonly employed characterization tools, can give rise to extrinsic magnetic glassiness [7]. Our earlier work, thus, highlights the importance of characterizing phase-pure samples for highly sensitive magnetization studies. During our past efforts in establishing the elusive magnetic ordering in Mn-doped SrTiO$_3$ samples, we had naturally attempted converting insulating SrTiO$_3$ to a metallic state via doping of electrons attained through oxygen vacancies [8], since the presence of charge carriers are known to give rise to ferromagnetic ordering by promoting magnetic interactions between localized magnetic moments by several different mechanisms, such as double exchange [9], RKKY [10-12] and kinetic stability [13]. All our phase-pure metallic samples of Mn doped SrTiO$_{3-\delta}$ invariably exhibited only a paramagnetic behavior without any evidence of any M-H loop down to 10 K. This negative result attains particular significance in view of a recent claim in the literature [14] that Mn-doped metallic SrTiO$_3$ shows long-range magnetic ordering due to RKKY interaction. The importance of such a study to ascertain the true nature of magnetism in similar dilute magnetic semiconductors has been exemplified earlier in the literature in the face of claims of intrinsic magnetism in Mn-doped ZnO [15].

**Results**

We have already reported our synthesis of SrTi$_{1-x}$Mn$_x$O$_{3.0}$ with $x$=0.02 and its properties in our earlier publication [7]. In order to dope electrons as charge carriers in the system, we introduce oxygen vacancies in the sample, forming SrTi$_{1-x}$Mn$_x$O$_{3-\delta}$. This is achieved by mixing SrTi$_{1-x}$Mn$_x$O$_{3.0}$ [7] and SrTiO$_{2.80}$ [8] in required proportions to achieve targeted $x$ and $\delta$ values and by repeated melting of the mixture in an arc furnace under high purity argon atmosphere. Additionally, we carried out thermogravimetric analysis (TGA) to quantify the oxygen content in each case, thereby defining the chemical formula of the samples. This way, two specific oxygen deficient samples, namely SrTi$_{0.98}$Mn$_{0.02}$O$_{2.99}$ and SrTi$_{0.99}$Mn$_{0.01}$O$_{2.96}$, were prepared for the present investigation. Room temperature X-ray diffraction of all three samples (Fig. 1a) establish phase pure cubic $P$m-3m space group. Fig. 1b shows the XRD data for all samples, plotted in a log scale that accentuates the small signals from impurity phases, if present, as was shown in our earlier publication [7]. We specifically expand the region between 30$^0$ and 40$^0$ in the inset of Fig. 1b, where the magnetic impuitry phase arising from Mn$_3$O$_4$ is known to show a diffraction peak [7]; clearly, there is no trace of this magnetic impurity phase in any of our samples. SrTi$_{0.98}$Mn$_{0.02}$O$_{2.99}$ sample was surprisingly found to be highly insulating in spite of the electron doping via oxygen vacancies; this highly insulating nature precluded any temperature dependent resistivity measurement on this sample, with the room temperature resistivity being larger than 10$^6$ ohm-cm. SrTi$_{0.99}$Mn$_{0.01}$O$_{2.96}$ sample was found to be metallic, as evidenced by the plot of the resistivity as a function of the temperature, T, in Fig.2. In order to understand the insulating nature of one sample and metallic property of the other, in spite of both samples being doped with oxygen vacancies, we have investigated the state of Mn using electron paramagnetic resonance (EPR).

Fig.3 shows the EPR spectra of SrTi$_{0.98}$Mn$_{0.02}$O$_{2.99}$ and SrTi$_{0.99}$Mn$_{0.01}$O$_{2.96}$ compared to that of SrTi$_{0.98}$Mn$_{0.02}$O$_{3.0}$, taken from ref.[7]. As discussed in ref.[7,16], the EPR spectrum of SrTi$_{0.98}$Mn$_{0.02}$O$_{3.0}$ sample in absence of any oxygen vacancy exhibits Landé g-factor, $g$=1.997 and a hyperfine splitting of 74 Oe, characteristic of Mn$^{4+}$ in an octahedral [Ti] site. The sample with the same level of Mn doping, but with an oxygen vacancy, $\delta$, equal to 0.01, clearly shows two sets of sextet lines, one set corresponding to $Mn^{4+}$ ions in octahedral sites with a hyperfine splitting of 74 Oe. The other set of sextet structure has a Landé $g$-factor of 2.005 and a hyperfine splitting of 83.5 Oe, suggesting $Mn^{2+}$ species at the octahedral site [3]. Combined with the insulating nature, this clearly suggests that doped electrons due to oxygen vacancies are effectively trapped by $Mn^{4+}$ sites, converting $Mn^{4+}$ to $Mn^{2+}$ and retaining the insulating nature of the sample. Thus, it suggests that we need to have oxygen vacancy at least comparable to Mn doping (i.e. $\delta \approx x$) for a sample of $SrTi_{1-x}Mn_xO_{3-\delta}$ to be metallic after converting all $Mn^{4+}$ to $Mn^{2+}$ species. This interpretation is validated by the observation of $SrTi_{0.99}Mn_{0.01}O_{2.96}$ sample being metallic. Additionally, the EPR spectrum of this sample shows exclusive presence of only $Mn^{2+}$ ions in octahedral sites (see Fig.3).

We have investigated the possibility of a magnetic order in our phase pure, high quality samples by carrying out magnetization ($M$) vs. applied magnetic field ($B$) at a low temperature of 10 K. The results for $M(B)$ of $SrTi_{0.98}Mn_{0.02}O_{2.99}$ and $SrTi_{0.99}Mn_{0.01}O_{2.96}$ are plotted in Fig.4, as we have already reported the properties of $SrTi_{0.98}Mn_{0.02}O_{3.0}$ as paramagnetic in ref.[7]. Fig.4 makes the absence of spontaneous magnetic order in either of the samples evident by the absence of any hysteresis in the $M$-$B$ plot. We have analyzed the $M$-$B$ plot in terms of a Brillouin function fit to the lineshape, since the magnetic response of a paramagnetic system is described by a Brillouin function. The solid lines through the data points in Fig.4, representing the best fits in each case, describe the experimental data very well. In the case of insulating $SrTi_{0.98}Mn_{0.02}O_{2.99}$, the magnetization data in Fig.4a required two Brillouin functions, one for $Mn^{2+}$ with $J = 5/2$ and the other for $Mn^{4+}$ with $J = 3/2$ spin-only values [17] with abundances of $Mn^{2+}$ and $Mn^{4+}$ in one unit cell equal to 0.003 and 0.009, respectively, thereby giving rise to an average moment of 4.4 $\mu_B$. The relative abundances of $Mn^{2+}$ and $Mn^{4+}$ are qualitatively consistent with the EPR spectrum of this compound in Fig.3b, showing the presence of both species. The $Mn^{2+}$ and $Mn^{4+}$ contents per unit cell are also consistent with the average Mn composition of the sample within the experimental and synthetic uncertainties. The magnetization data of the metallic $SrTi_{0.99}Mn_{0.01}O_{2.96}$ could be fitted with a single Brillouin function corresponding to $Mn^{2+} J = 5/2$ state with the $Mn^{2+}$ content being 0.005 per unit cell, once again consistent with the EPR data (Fig.3c) and the composition. The effective magnetic moment of $SrTi_{0.99}Mn_{0.01}O_{2.96}$ extracted from the slope of the low temperature Curie-Weiss plot is 3.8 $\mu_B$ per Mn ion, which is close to the expected value of 5.9 $\mu_B$ for ionic $Mn^{2+}$ ions within the aforementioned uncertainties in determining the exact composition of dilute constituents, illustrating paramagnetic contribution from nearly all Mn ions in the sample. Interestingly, both metallic and insulating samples show a small negative intercept (in the range of ~ -5 K) of the low temperature Curie-Weiss plot, suggesting only a weak antiferromagnetic interaction between the Mn sites.

**Discusssion**

The EPR spectra in Fig.3 show dominantly the narrow sextet pattern of hyperfine splitting of isolated Mn ions, doped at octahedral Ti site, for both insulating and metallic samples, with negligible contribution from any broad Mn signal. Broad Mn signal in EPR is known to arise from Mn-Mn dipolar interaction and is indicative of clustering of Mn- ions. In our earlier studies [7], we found a correlation between a significant presence of such a dipolar broadened EPR spectrum and the occurrence of extrinsic ferromagnetic signal in the system. It is important to note here that the insulating $SrTi_{0.95}Mn_{0.05}O_3$ (termed OSTMN) of ref.[14] exhibiting a sharp sextet signal is paramagnetic, but the same sample, somewhat reduced, to form a metallic $SrTi_{0.95}Mn_{0.05}O_{3-\delta}$ exhibits both a strong dipole broadened EPR component and ferromagnetism.

The Brillouin function analysis could very nicely explain the magnetization data of both insulating $SrTi_{0.98}Mn_{0.02}O_{2.99}$ and metallic $SrTi_{0.99}Mn_{0.01}O_{2.96}$ samples, as shown in Fig. 4, and the extracted paramagnetic Mn dopings from the fits to the experimental data were found to be consistent with corresponding nominal Mn doping levels, as discussed in the earlier section. This internal consistency check may be contrasted against the results presented in ref.[14], claiming ferromagnetism in their nominally $SrTi_{0.95}Mn_{0.05}O_{2.85}$ sample. The saturation magnetization of the M-H loop reported there corresponds to a moment of ~0.04 $\mu_B$/ Mn ion [or 0.002 $\mu_B$ per unit cell]; in other words, roughly one Mn ion out of a hundred doped ones contributes to the ferromagnetic signal in that sample. Even if we consider a significant loss in Mn content in the reported sample and compare the paramagnetic moment from the slope of $\chi^{-1}$ vs. T plot at the high temperature of that sample to estimate the actual Mn content, we find that the unit cell has about 0.03 $Mn^{2+}$ ions. Even with this correction, it would appear that only a small fraction (~ 1%) of the total Mn content gives rise to the ferromagnetic signal reported in ref.[14], while the remaining 99% of Mn presumably remains paramagnetic. This interpretation, consistent with the absence of any signature of a ferromagnetic transition in the *M* vs. *T* plot, together with the observed dipole broadened EPR signal only in the metallic sample, points to an extrinsic origin of the ferromagnetic signal in that report.

In conclusion, we have synthesized electron and Mn doped $SrTiO_3$ samples, doped through generation of varying amounts of oxygen vacancies in these samples. We find that for a small oxygen vacancy doping, doped electrons are trapped by $Mn^{4+}$ sites, converting them partially to $Mn^{2+}$, while the sample remained a strong insulator. However, for larger oxygen vacancy doping, namely $SrTi_{0.99}Mn_{0.01}O_{2.96}$, the sample becomes metallic, accompanied by a complete reduction of $Mn^{4+}$ ions to $Mn^{2+}$. We find that our phase pure Mn doped $SrTiO_3$ samples, both in insulating and metallic ground states, are paramagnetic down to the lowest measurement temperatures. Unlike recent literature reporting weak ferromagnetism in metallic Mn doped $SrTiO_3$ samples, we do not find any enhancement of magnetic couplings in the metallic sample compared to the insulating samples.

**Methods**

To achieve targeted electron and Mn doping levels into $SrTiO_3$, stoichiometric amounts of $SrTi_{1-x}Mn_xO_{3.0}$ [7] and $SrTiO_{2.80}$ were mixed thoroughly and arc melted repeatedly in high-purity Ar gas atmosphere. $SrTiO_{2.80}$ was prepared by arc melting stoichiometric amounts of $Sr_2TiO_4$, $TiO_2$ and Ti [8]. EPR experiments were performed at 9.8 GHz in a ER 200D X-band Bruker Spectrometer. Resistivity measurements were performed using van-der Pauw geometry [18]. TGA experiments were performed by heating the sample in air and monitoring the weight gain on complete oxidation. Magnetization measurements were performed in a Quantum-design SQUID magnetometer.

**Acknowledgments**

Authors thank Department of Science and Technology, Board of Research in Nuclear Sciences, and Centre for Scientific and Industrial Research, Government of India for funding and support.

**Figure Legends**

Fig 1: XRD data of SrTi$_{0.98}$Mn$_{0.02}$O$_{3.0}$ (O$_{3.0}$), SrTi$_{0.98}$Mn$_{0.02}$O$_{2.99}$ (O$_{2.99}$) and SrTi$_{0.99}$Mn$_{0.01}$O$_{2.96}$ (O$_{2.96}$) plotted in linear scale (a) and in log scale (b) to illustrate the phase purity of the synthesized samples. The inset of Fig 1(b) highlights the absence of any Mn$_3$O$_4$ related magnetic impurity phase, which is expected to have a maximum XRD peak around 2θ=36°.

Fig 2: Resistivity of SrTi$_{0.99}$Mn$_{0.01}$O$_{2.96}$ showing it to be metallic down to the lowest measurement temperature.

Fig 3: Electron Paramagnetic Resonance spectra of SrTi$_{0.98}$Mn$_{0.02}$O$_{3.0}$ (a), SrTi$_{0.98}$Mn$_{0.02}$O$_{2.99}$ (b) and SrTi$_{0.99}$Mn$_{0.01}$O$_{2.96}$ (c). The spectrum (a) is taken from ref.[7].

Fig 4: Magnetization of $SrTi_{0.98}Mn_{0.02}O_{2.99}$ (a) and $SrTi_{0.99}Mn_{0.01}O_{2.96}$ (b) collected at 10 K. The data are shown by symbols. The Brillouin function fits are shown by solid lines running through the corresponding data.

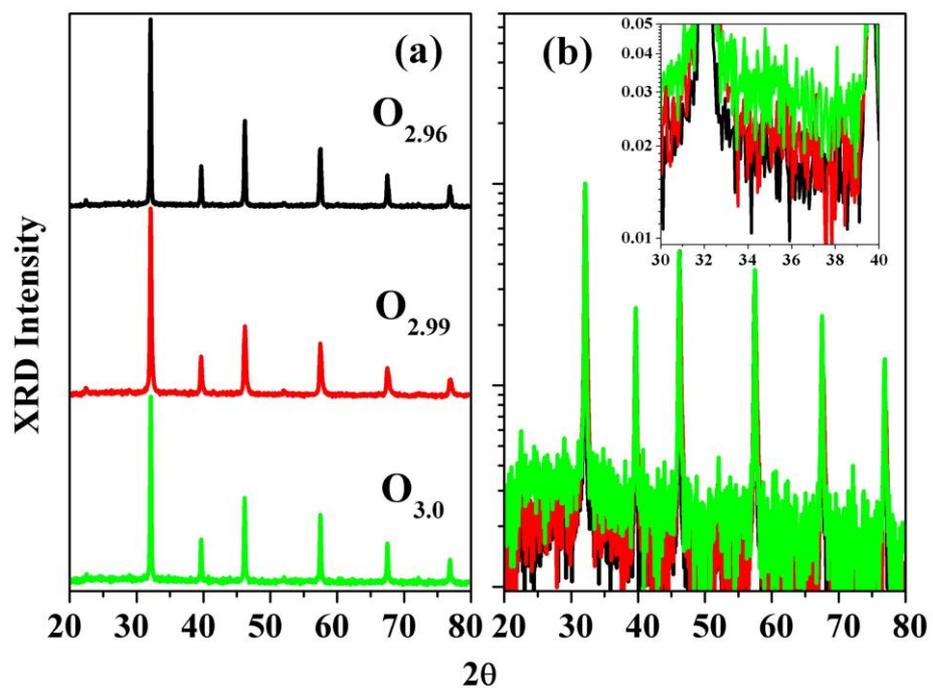

Fig. 1

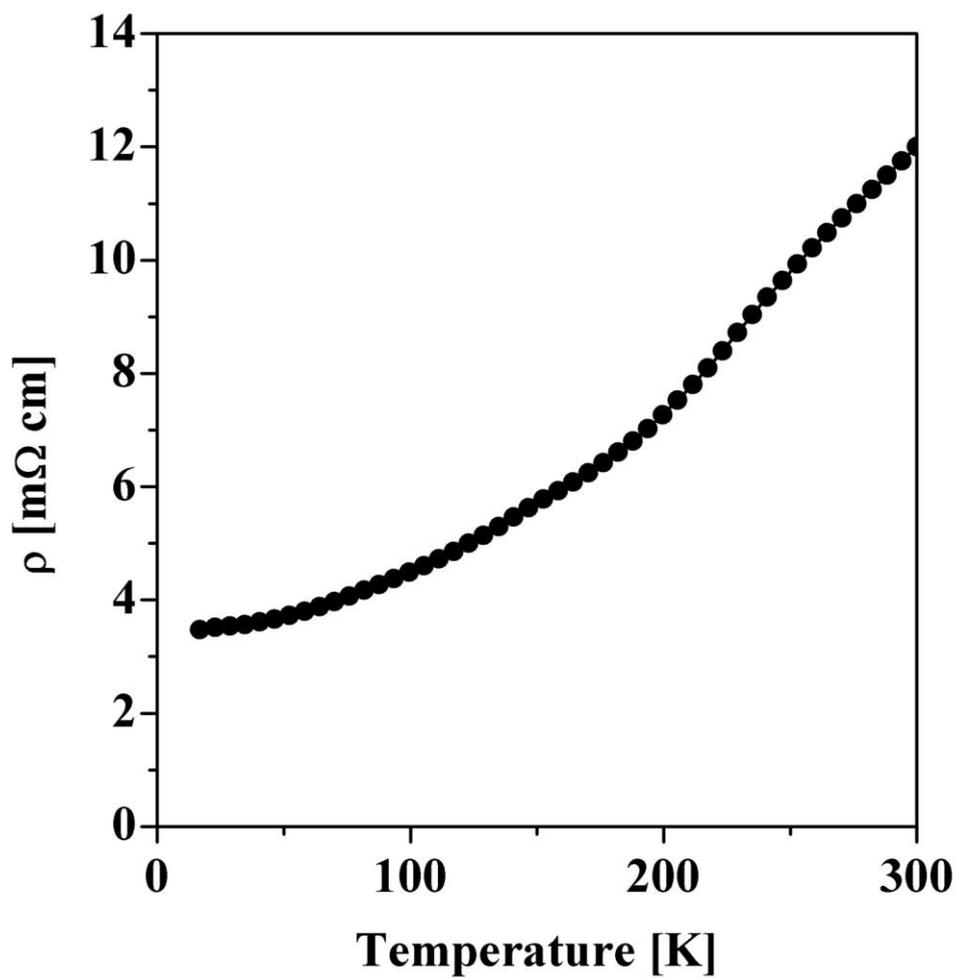

**Fig. 2**

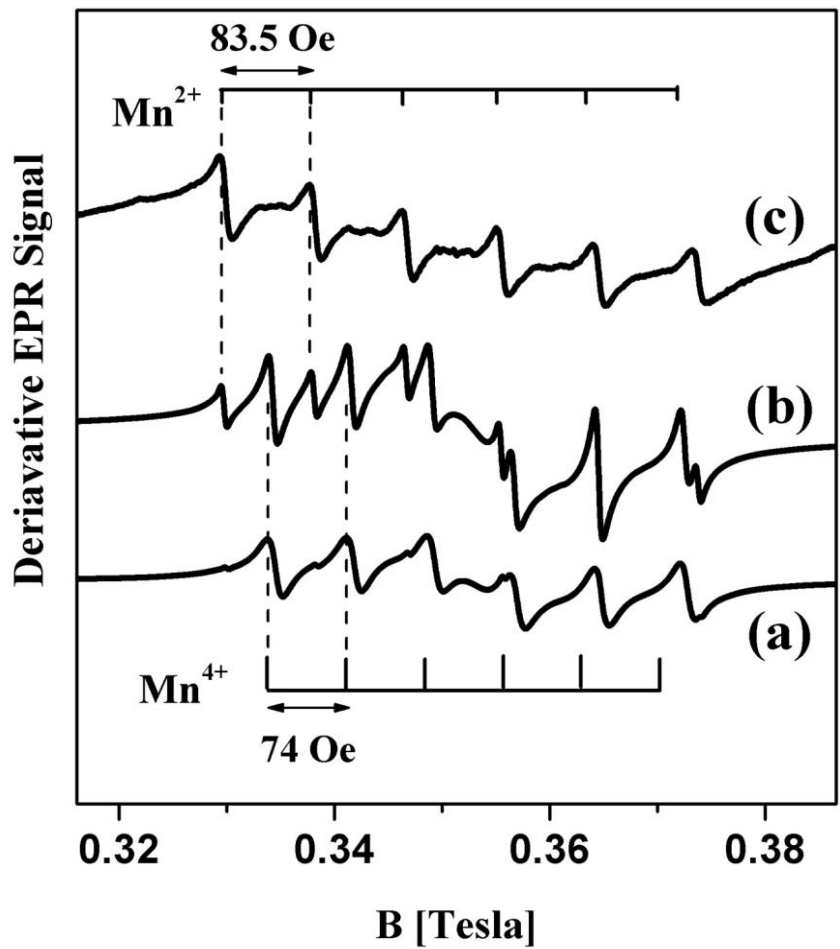

Fig. 3

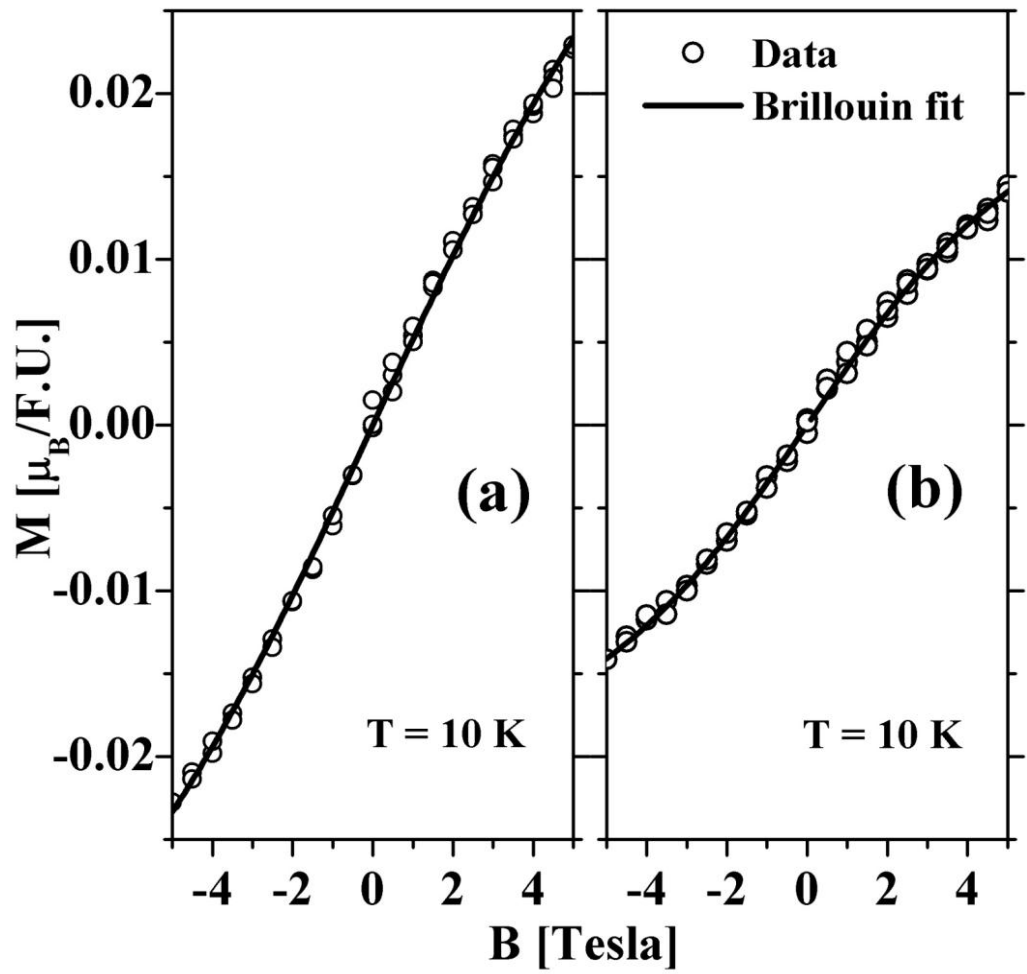

Fig. 4